\documentclass[11pt]{article}
\usepackage{graphicx}

\begin{document}

\title{\textbf{\textsf{Cosmic coincidence problem and variable constants of physics}}}
\author{Mubasher Jamil\footnote{mjamil@camp.edu.pk},
Farook Rahaman\footnote{farook\_rahaman@yahoo.com}, and Mehedi Kalam\footnote{mehedikalam@yahoo.co.in}\\ \\
$^\ast$\small Center for Advanced Mathematics and Physics, National
University of Sciences and Technology,\\ \small Peshawar Road,
Rawalpindi - 46000, Pakistan\\
$^\dag$\small Department of Mathematics, Jadavpur University,
Kolkata - 700032, India
\\
$^\ddag$\small Department of Physics, Netaji Nagar College for
Women, Regent Estate, Kolkata - 700092, India
 }\maketitle
\begin{abstract}
The standard model of cosmology is investigated using time-dependent
cosmological constant $\Lambda$ and Newton gravitational constant
$G$. The total energy content is described by the modified Chaplygin
gas equation of state. It is found that the time-dependent constants
coupled with the modified Chaplygin gas interpolate between the
earlier matter to the later dark energy dominated phase of the
universe. We also achieve a convergence of the parameter
$\omega\rightarrow-1$, almost at the present time. Thus our model
fairly alleviates the cosmic-coincidence problem which demands
$\omega=-1$ at the present time.
\end{abstract}

\newpage
\large
\section{Introduction}
The astrophysical observations of supernovae of type Ia give a
convincing evidence of a universe undergoing an accelerated phase of
its expansion history; rather than going to deceleration as was
expected theoretically \cite{perl,riess,tonry,wang}. Similar
conclusion has been deduced from the observations of anisotropies in
cosmic microwave background by the WMAP \cite{spergel,bennett} which
favors a low density, spatially flat ($\Omega_{tot}\sim1$) universe
filled with an exotic vacuum energy containing the maximum portion
of the total energy density, $\Omega_{\Lambda}\sim0.7$ \cite{perl1}.
This mysterious dark energy is represented by a barotropic equation
of state (EoS) $p=\omega\rho$, where $p$ and $\rho$ is the pressure
and the energy density of dark energy, with $\omega\leq-1$. If
$\omega=-1$, it is called `cosmological constant', $-1<\omega<-1/3$
is dubbed `quintessence', while it is called `phantom energy' if
$\omega<-1$ \cite{caldwell,caldwell1}. All these candidates have
some fundamental problems: The former one presents discrepancy of
$120$ orders of magnitude between theoretical and empirical results
\cite{sahni}, while quintessence requires fine tuning of
cosmological parameters for a suitable choice of the potential
function \cite{zlatev}. Also the phantom energy yields eccentric
predictions like `big rip' and ripping apart of gravitationally
bound objects \cite{babichev,jamil33}. Moreover, the variations in
$\omega$ suggest that there is no consensus on the actual EoS of
dark energy and one can only deal with the upper and lower bounds on
$\omega$ \cite{upadhye}.

Although the dark energy is generically considered to be a perfect
fluid yet it need not to be perfect if it experiences perturbations
\cite{fabris}. Moreover, the dark energy also may not be completely
`dark' especially if it gets coupled with matter and energy exchange
takes place; consequently the matter evolution becomes modified
\cite{linder1,nojiri}. It is now obvious that this sudden transition
to acceleration from the earlier deceleration phase is rather recent
with the corresponding redshift $z\leq1$ \cite{daly}. It has been
proposed that the standard model of cosmology may not be sufficient
to explain this exotic phenomenon and hence significant
modifications are proposed like a modified Friedmann equation
$H^2\sim g(\rho)$, where $g$ is an arbitrary function of $\rho$ and
$H$ is the Hubble parameter \cite{gong} and adding a Cardassian term
in the Friedmann equation \cite{freese}. Other possible explanations
proposed are dark energy arising from tachyonic matter
\cite{padmanabhan,diaz}, van der Waals fluid \cite{kremer},
geometric dark energy \cite{linder}, Dvali-Gabadadze-Porrati (DGP)
gravity model \cite{dvali} and Randall-Sundrum brane world model
\cite{randall} are most prominent. Although the nature of dark
energy is not clear but its thermodynamical properties suggest a
universe filled with it becomes hotter with time \cite{lima}.

One of the most interesting problems in the present cosmology is the
cosmic coincidence (or cosmic conundrum) problem which naively asks:
why the energy densities of matter and dark energy are of the same
order or the corresponding dimensionless ratio is closer to 1, at
current time \cite{campo,dalal,dodelson}. This problem can be posed
in terms of the EoS parameter $\omega$: if the parameter
$\omega\geq-1$ in the past and $\omega<-1$ in the future then why we
are observing $\omega=-1$ at present time. In recent years, this
problem is addressed using a notion of interacting dark energy model
in which both the interacting components i.e. dark energy and matter
exchange energy to keep the density ratio close to 1. This model has
some intrinsic problems that are still unresolved: the microphysics
of energy transfer is not exactly understood i.e. the particles that
can mediate the interaction are not pointed out. Moreover, the
coupling function (or the decay rate) for the required interaction
is chosen quite arbitrarily \cite{quartin,jamil,jamil1,jamil3} and
also the coupling constant involved is not yet properly constrained
theoretically or observationally \cite{jamil2}. In this paper, we
address this problem using a simplified approach by considering the
constants of physics to evolve over cosmic time. We here take three
ansatz for scale factor and analyze the behavior of parameter
$\omega$. Curiously, the parameter $\omega(z)$ evolves from positive
to negative values and finally converges to $-1$. This result turns
out to be consistent with the observations. Hence our model fairly
addresses the cosmic coincidence problem and practically alleviates
it.

Due to multitude of uncertainties in the determination of $\omega$
observationally and other intrinsic theoretical problems (as
discussed above) with it, we here proceed with an EoS commonly
called the Chaplygin gas (CG) represented by \cite{dev,kamen}
\begin{equation}
p=-\frac{X}{\rho},
\end{equation}
where $X$ is a constant parameter. The CG effectively explains the
evolution of the universe from the earlier deceleration (matter
dominated era) to the later acceleration phase (dark energy
dominated) as is manifested in the following equation
\begin{equation}
\rho=\sqrt{X+\frac{Y}{a^6}}.
\end{equation}
Here $Y$ is a constant of integration parameter. For small $a$, it
gives $\rho\sim\sqrt{Y}a^{-3}$ while for large $a$, we have
$\rho\sim\sqrt{X}$. Therefore models based on CG are also called
dark energy-matter unification models \cite{bilic,bento1}. Due to
its this effectiveness, several generalizations of CG are proposed
(see e.g. \cite{surajit,debnath,writ,setare,setare1,guo,sen,zhang}).
The Chaplygin gas arises from the dynamics of a generalized
$d$-brane in a ($d+1,1$) spacetime and can be described by a complex
scalar field whose action can be written as a generalized
Born-Infeld action \cite{bento}.

The plan of the paper is as follows: In the second section, we shall
present the model of our system. In third section, we determine the
cosmological parameters for different choices of the scale factor
parameter. The last section is devoted for the conclusion of our
paper.

\section{The cosmological model with variable constants}

The Friedmann-Robertson-Walker (FRW) metric which satisfies the
cosmological principle is specified by
\begin{equation}
ds^2=-dt^2+a^2(t)\left[
\frac{dr^2}{1-kr^2}+r^2(d\theta^2+\sin^2\theta d\phi^2) \right].
\end{equation}
Here $a(t)$ is the scale factor that determines the expansion of the
universe. Also the parameter $k$ is the curvature parameter
determining the spatial geometry of the FRW spacetime. It can take
three possible values $k=+1,0,-1$ which correspond to spatially
closed, flat and open universe respectively or geometrically
spherical, Minkowskian and hyperbolic spacetime respectively.

The equations of motion corresponding to FRW metric are
\begin{eqnarray}
H^2\equiv\left(\frac{\dot{a}}{a}\right)^2&=&\frac{8\pi G}{3}\rho+\frac{\Lambda}{3},\\
\frac{\ddot{a}}{a}&=&-\frac{4\pi G}{3}(\rho+3p)+\frac{\Lambda}{3}.
\end{eqnarray}
Above $H$ is the Hubble parameter. Note that we have assumed $k=0$
in the above equations which is favored by the observational data.
Here $\Lambda$ is called the cosmological constant with dimensions
of $(length)^{-2}$. Note that Eq. (5) shows that accelerated
expansion of the universe $\ddot{a}>0$ is possible if the strong
energy condition $\rho+3p>0$ is violated and also it is independent
of the choice of $k$. The energy conservation equation for the above
system is
\begin{equation}
\dot{\rho}+3H(\rho+p)=0.
\end{equation}
The cherished constants of physics that describe the universe need
not to be constant but can vary with respect to other parameters.
For instance, the cosmological constant
$\Lambda(t)=\Lambda(t_o)+(t-t_o)\dot{\Lambda}(t_o)+...$ which is
constant at zeroth order approximation but is really a time
dependent function at higher order approximations. Note that
accelerated expansion of the universe follows from
$\dot{\Lambda}>0$. The cosmic history of $\Lambda$ shows that it was
large in the past while it is small at present and will continue to
decrease, hence it gives a parametrization $\Lambda\propto
t^\sigma$, $\Lambda\propto\rho^\gamma$ and $\Lambda\propto H^2$
\cite{mukho,fomin}. A variable cosmological constant also arises in
theories of higher spatial dimensions like string theory and
manifests itself as the energy density for the vacuum \cite{hongya}
and it can also addresses the cosmic age problem effectively
\cite{ray}. Similarly, there is some evidence of a varying Newton's
constant $G$: Observations of Hulse-Taylor binary pulsar B$1913+16$
gives a following estimate
$0<\dot{G}/G\sim2\pm4\times10^{-12}{yr}^{-1}$ \cite{kogan},
helioseismological data gives the bound
$0<\dot{G}/G\sim1.6\times10^{-12}{yr}^{-1}$ \cite{guenther} (see Ref
\cite{ray1} for various bounds on $\dot{G}/G$ from observational
data). The variability in $G$ results in the emission of
gravitational waves. Dimensional analysis also shows that the time
dependent parameter $\Lambda$ to be decreasing with time $t$
\cite{jose,jose1}. In another approach, it is shown that $G$ can be
oscillatory with time \cite{pradhan}. It is recently proposed that
variable cosmic constants are coupled to each other i.e. variation
in one leads to changes in others \cite{vishwakarma}. A variable
gravitational constant also explains the dark matter problem as well
\cite{goldman}. Also discrepancies in the value of Hubble parameter
can be removed with the consideration of variable $G$
\cite{bertolami1}. Due to these reasons, we shall take $\Lambda$ and
$G$ to be time dependent quantities i.e. $\Lambda=\Lambda(t)$ and
$G=G(t)$. Hence Eqs. (4) and (5) yield
\begin{equation}
G\dot{\rho}+\rho\dot{G}+3(\rho+p)GH+\frac{\dot{\Lambda}}{8\pi}=0.
\end{equation}
Using Eqs. (6) and (7), we can write
\begin{equation}
\rho\dot{G}+\frac{\dot{\Lambda}}{8\pi}=0.
\end{equation}
Taking the ansatz for cosmological constant as \cite{arbab}
\begin{equation}
\Lambda=\frac{3\beta}{\rho^\gamma},
\end{equation}
where $\beta$ and $\gamma$ are constant parameters. Note that this
is a general ansatz and can reduce to Chakraborty and Debnath
\cite{chakraborty} if $\gamma=-1$. Using the modified Chaplygin gas
(MCG) EoS given by \cite{jing}
\begin{equation}
p=A\rho-\frac{B}{\rho^\alpha},
\end{equation}
where $A$ and $B$ are constant parameters and $0\leq\alpha\leq1$.
Thermodynamical analysis of MCG show that the values $\alpha=1/4$
and $B=1/3$ are consistent with the phenomenological results
\cite{bedran}. It is also shown that the recent supernovae data
favors $\alpha>1$ values \cite{bertolami,bento2}. The MCG reduces to
generalized Chaplygin gas (GCG) if $A=0$ while it gives CG if
further $\alpha=1$. While a barotropic EoS is obtained if $B=0$.
Thus Eq. (10) is a combination of a barotropic and GCG EoS.
Precisely, the observations of cosmic microwave background gives the
constraint $-0.35\leq A\leq0.025$ at $95\%$ confidence level
\cite{liu}. Analysis of various cosmological models show that models
based on Chaplygin gas best fit with supernova data \cite{grigoris}.
Using Eq. (10) in (6), we get the density evolution of MCG as
\begin{equation}
\rho=\left( x+Ca^{-y} \right)^{\frac{1}{1+\alpha}},
\end{equation}
where $x=\frac{B}{1+A}$ and $y=3(1+\alpha)(1+A).$ Making use of Eqs.
(9) and (11) in (8), the parameter $G$ is determined to be
\begin{equation}
G=\frac{-3\beta\gamma C}{8\pi (\alpha+\gamma+2)}\left[
(x+Ca^{-y})^{-\delta}\left(1+\frac{xa^y}{C}\right)^\delta
{_{2}F_{1}}\left(\delta,\delta,1+\delta,-\frac{xa^y}{C} \right)
\right],
\end{equation}
where
\begin{equation}
\delta=\frac{\alpha+\gamma+2}{1+\alpha}.
\end{equation}
The parameter $\Lambda$ can alternatively be written as
\begin{equation}
\Lambda=3\beta(x+Ca^{-y})^{\frac{-\gamma}{1+\alpha}}.
\end{equation}

\section{Determination of cosmological parameters}
To analyze the behavior of the above cosmological parameters, we
will consider three cases:
\begin{enumerate}
  \item $a\simeq a_0 T^n,$
  \item $a\simeq(uT - v)^{\frac{1}{(1+q)}}$,
  \item $a\simeq[ e^{-DPT} - 1 ]^{-\frac{1}{P}}.$
\end{enumerate}
Here $D$, $P$, $q$, $u$, $v$, $n$ and $a_o$ are constant parameters.
Also $T=t/t_o$ is the dimensionless time parameter with $t_o$ is the
current age of the universe.

\subsection{Power law form of scale factor}

We consider power law form of the scale factor
\begin{equation}
a\simeq a_o T^n,
\end{equation}
where $a_o$ and $n$ are arbitrary constants. For this choice, it is
possible to get the accelerated expansion of the universe if $n>1$.
Now, all the physical parameters will take the following forms as:
\begin{eqnarray}
 \rho&\simeq&\left( x+\frac{C}{(a_oT^n)^{y}} \right)^{\frac{1}{1+\alpha}},\\
G&\simeq&\frac{-3\beta\gamma C}{8\pi (\alpha+\gamma+2)}[
\left(x+\frac{C}{(a_oT^n)^{y}}\right)^{-\delta}\left(1+\frac{x(a_oT^n)^{y}}{C}\right)^\delta\nonumber\\
&\;&\times
{_{2}F_{1}}\left(\delta,\delta,1+\delta,-\frac{x(a_0T^n)^{y}}{C}
\right)],\\
\Lambda&\simeq&3\beta\left(x+\frac{C}{(a_oT^n)^{y}}\right)^{\frac{-\gamma}{1+\alpha}},\\
p&\simeq&A\left[\left( x+\frac{C}{(a_oT^n)^{y}}
\right)^{\frac{1}{1+\alpha}}\right]-\frac{B}{\left[\left(
x+\frac{C}{(a_oT^n)^{y}}
\right)^{\frac{1}{1+\alpha}}\right]^\alpha}.
\end{eqnarray}
The cosmological parameters obtained in this section are plotted in
figures 1 to 4 against parameter $T$.

\begin{figure}
\includegraphics{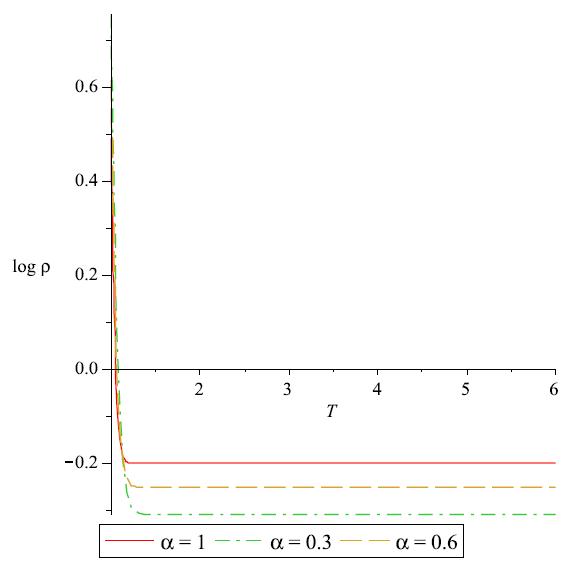}\\
\caption{The energy density $\rho$ (see Eq. 16) is plotted against
time $T$. Model parameters are fixed at $A=B=C=n=2$, $a_o=1$,
$\gamma=-0.5$ and $\beta=1$. }
\end{figure}
\begin{figure}
\includegraphics{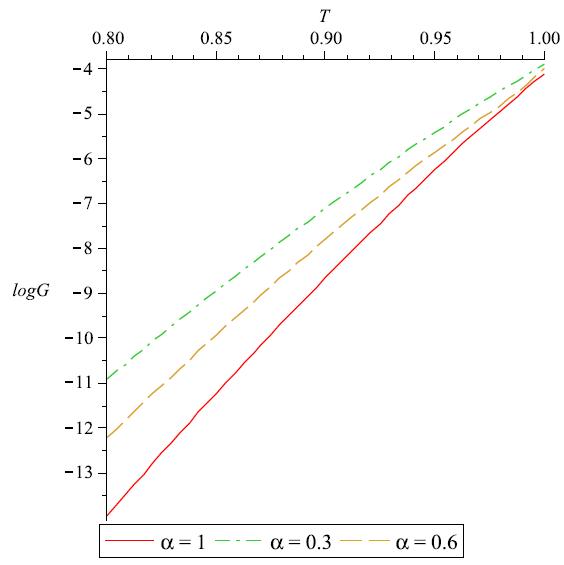}\\
\caption{The parameterized Newton's constant $G$ (see Eq. 17) is
plotted against time parameter $T$ for different choices of
$\alpha$. Model parameters are fixed as taken in Fig.1}
\end{figure}
\begin{figure}
\includegraphics{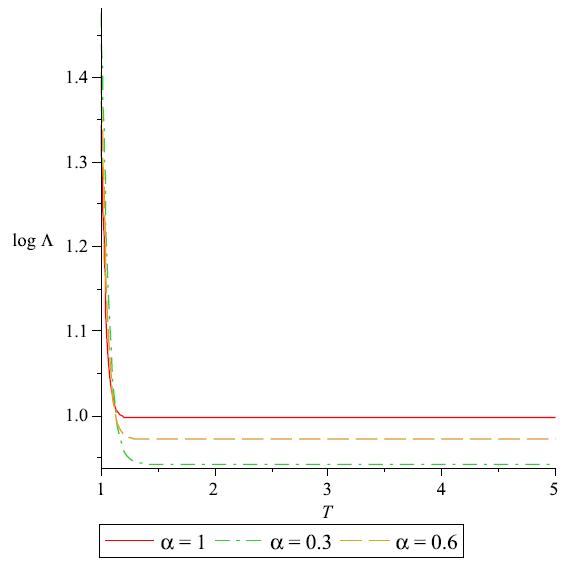}\\
\caption{The cosmological constant $\Lambda$ (see Eq. 18) is plotted
against time parameter $T$. Model parameters are fixed as taken in
Fig.1}
\end{figure}
\begin{figure}
\includegraphics{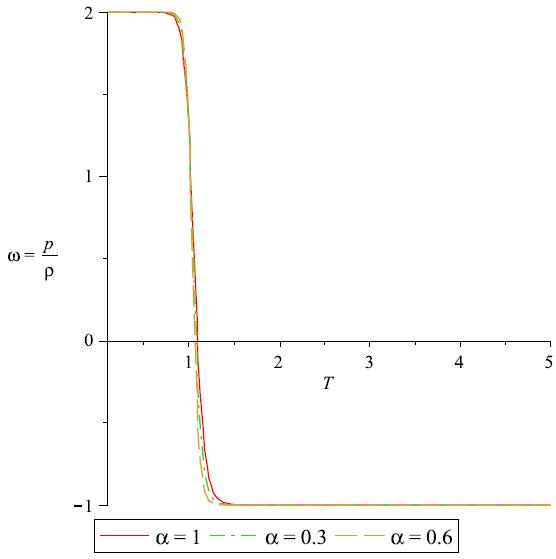}\\
\caption{The EoS parameter $\omega=p/\rho$ (see Eq. 16 and 19) is
plotted against the time parameter $T$. Model parameters are fixed
as taken in Fig.1}
\end{figure}
\newpage
\subsection{Negative constant deceleration parameter}
In this case, we consider constant deceleration parameter model
defined by
\begin{equation}
-\frac{\ddot{a}}{aH^2} = constant = q.
\end{equation}
Here the constant is taken to be negative i.e. it is an accelerating
model of the universe \cite{raha}. The solution of equation (20) is
\begin{equation}
 a \simeq ( uT - v)^{\frac{1}{(1+q)}},
\end{equation}
where, $u$ and $v$ are integration constants. This equation implies,
the condition of expansion is $1+q>0$. Now, all the physical
parameters will take the following forms as:
\begin{eqnarray}
\rho&\simeq&\left( x+\frac{C}{( uT - v)^{\frac{y}{(1+q)}}}
\right)^{\frac{1}{1+\alpha}},\\
G&\simeq&\frac{-3\beta\gamma C}{8\pi (\alpha+\gamma+2)}[
\left(x+\frac{C}{( uT -
v)^{\frac{y}{(1+q)}}}\right)^{-\delta}\left(1+x\frac{( uT -
v)^{\frac{y}{(1+q)}}}{C}\right)^\delta\nonumber\\
&\;&\times {_{2}F_{1}}\left(\delta,\delta,1+\delta,-x\frac{( uT -
v)^{\frac{y}{(1+q)}}}{C} \right)],\\
\Lambda&\simeq&3\beta\left(x+\frac{C}{( uT - v)^{\frac{y}{(1+q)}}}
\right)^{\frac{-\gamma}{1+\alpha}},\\
p&\simeq&A\left[\left( x+\frac{C}{( uT - v)^{\frac{y}{(1+q)}}}
\right)^{\frac{1}{1+\alpha}} \right]-\frac{B}{\left[\left(
x+\frac{C}{( uT - v)^{\frac{y}{(1+q)}}}
\right)^{\frac{1}{1+\alpha}}\right]^\alpha}.
\end{eqnarray}
The cosmological parameters obtained in this section are plotted in
figures 5 to 8 against parameter $T$.
\newpage

\begin{figure}
\includegraphics{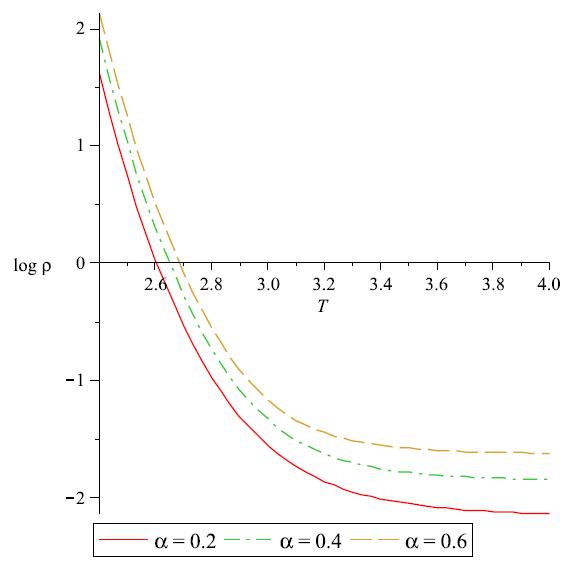}\\
\caption{The density parameter $\rho$ (see Eq. 22) is plotted
against time $T$. Model parameters are fixed at $A=B=C=0.08$, $u=1$,
$v=2$, $q=\gamma=-0.2$ and $\beta=1$.}
\end{figure}
\begin{figure}
\includegraphics{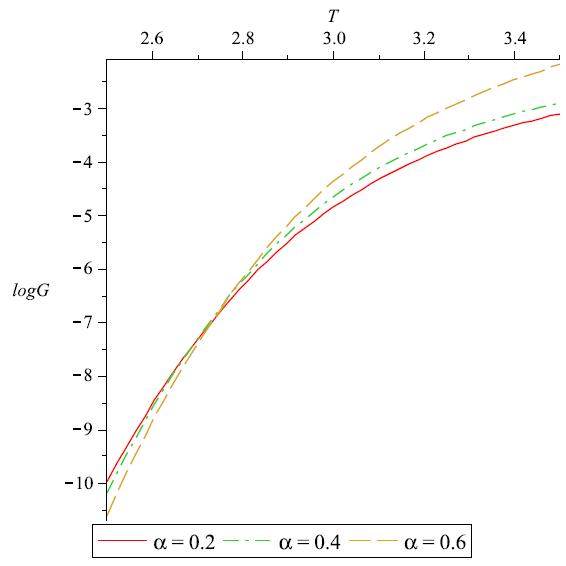}\\
\caption{The Newton's constant $G$ (see Eq. 23) is plotted against
time parameter $T$ for different choices of $\alpha$. Other model
parameters are fixed as in Fig.5}
\end{figure}
\begin{figure}
\includegraphics{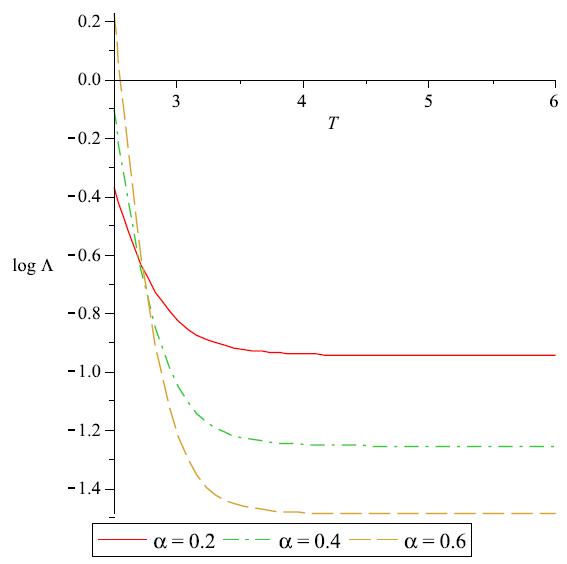}\\
\caption{The cosmological constant $\Lambda$ (see Eq. 24) is plotted
against time parameter $T$. Other model parameters are fixed as in
Fig.5}
\end{figure}
\begin{figure}
\includegraphics{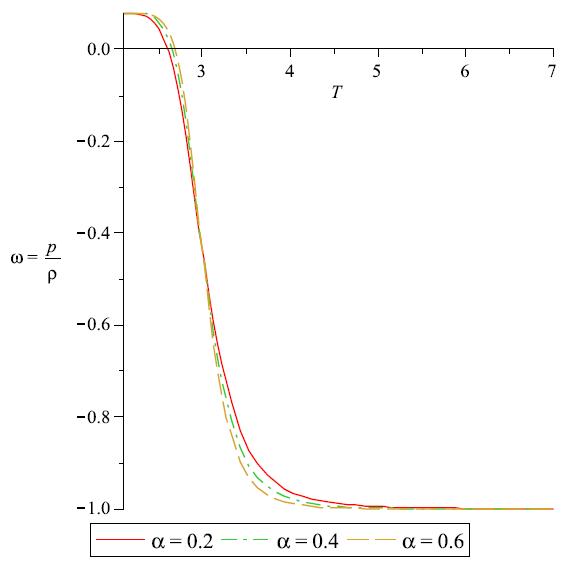}\\
\caption{The parameter $p/\rho=\omega$ (see Eq. 22 and 25) is
plotted against the time parameter $T$. Other model parameters are
fixed as in Fig.5}
\end{figure}

\pagebreak
\newpage

\subsection{Specific form of deceleration parameter}

In this case, we consider specific form  deceleration parameter
model defined by \cite{baner}
\begin{equation}
-\frac{\ddot{a}}{aH^2} = -1 - \frac{P a^P}{1 + a^P}.
\end{equation}
Here $P$ is is a constant. This choice of deceleration parameter
provides an early deceleration and late time acceleration of the
universe. The solution of equation (26) is
\begin{equation}
a \simeq [ e^{-DPT} - 1 ]^{-\frac{1}{P}},
\end{equation}
where $D$ is an  integration constant. For negative values of $P$,
we always get accelerated expansion of the universe. Now, all the
physical parameters will take the following forms as:
\begin{eqnarray}
\rho&\simeq&\left( x+\frac{C}{[ e^{-DPT} - 1 ]^{-\frac{y}{P}}}
\right)^{\frac{1}{1+\alpha}},\\
G&\simeq&\frac{-3\beta\gamma C}{8\pi (\alpha+\gamma+2)}[
\left(x+\frac{C}{[ e^{-DPT} - 1
]^{-\frac{y}{P}}}\right)^{-\delta}\left(1+x\frac{[ e^{-DPT} - 1
]^{-\frac{y}{P}}}{C}\right)^\delta\nonumber\\
&\;&\times {_{2}F_{1}}\left(\delta,\delta,1+\delta,-x\frac{[
e^{-DPT} - 1
]^{-\frac{y}{P}}}{C} \right)],\\
\Lambda&\simeq&3\beta \left( x+\frac{C}{[ e^{-DPT} - 1
]^{-\frac{y}{P}}}
\right)^{\frac{-\gamma}{1+\alpha}},\\
p&\simeq&A\left( x+\frac{C}{[ e^{-DPT} - 1 ]^{-\frac{y}{P}}}
\right)^{\frac{1}{1+\alpha}}-\frac{B}{\left[\left( x+\frac{C}{[
e^{-DPT} - 1 ]^{-\frac{y}{P}}}
\right)^{\frac{1}{1+\alpha}}\right]^\alpha}.
\end{eqnarray}
The cosmological parameters obtained in this section are plotted in
figures 9 to 12 against parameter $T$. \pagebreak
\newpage
\begin{figure}
\includegraphics{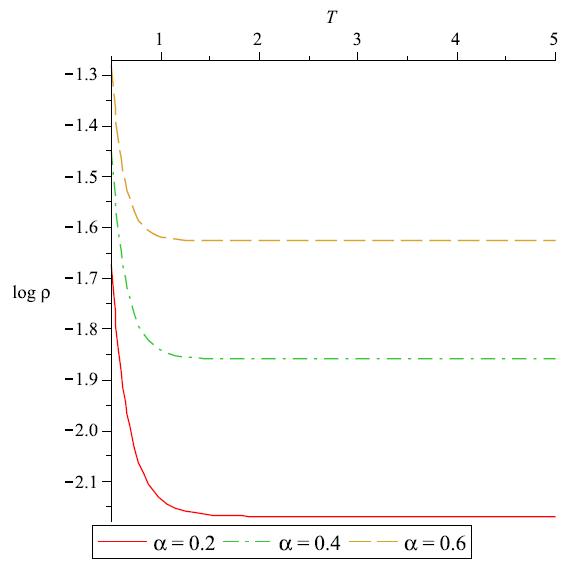}\\
\caption{The density parameter $\rho$ (see Eq. 28) is plotted
against time $T$. Model parameters are fixed at $A=B=C=0.08$,
$P=1.5$, $D=1$ and $\gamma=-0.2$}
\end{figure}
\begin{figure}
\includegraphics{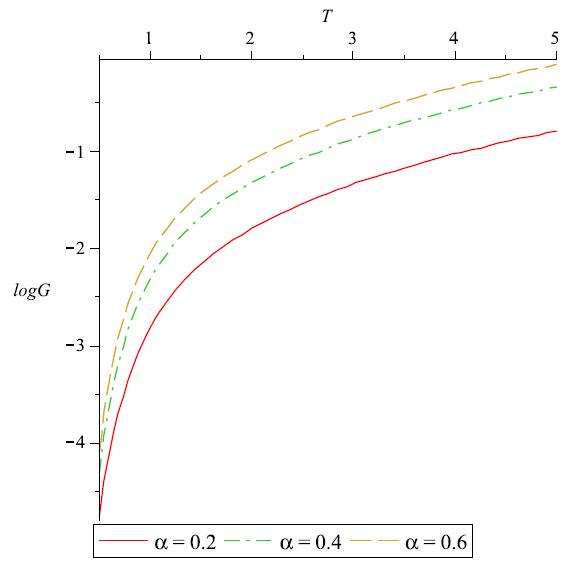}\\
\caption{The Newton's constant $G$ (see Eq. 29) is plotted against
time parameter $T$ for different choices of $\alpha$. Other model
parameters are fixed as in Fig.9}
\end{figure}
\begin{figure}
\includegraphics{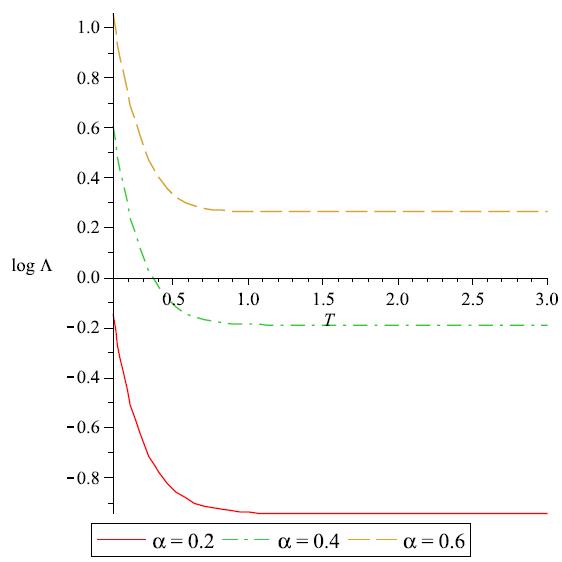}\\
\caption{The cosmological constant $\Lambda$ (see Eq. 30) is plotted
against time parameter $T$. Other model parameters are fixed as in
Fig.9}
\end{figure}
\begin{figure}
\includegraphics{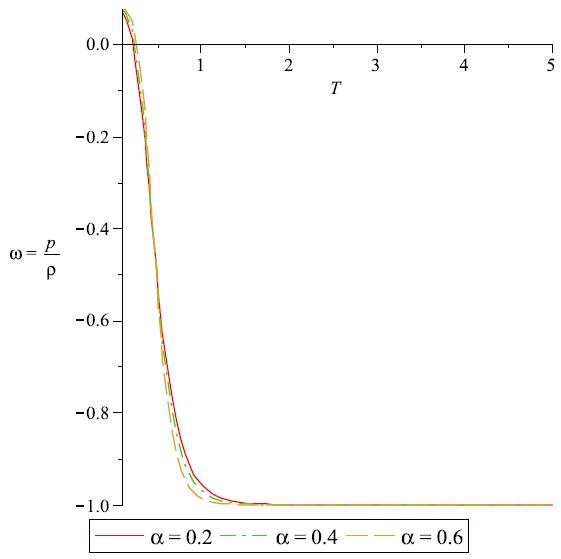}\\
\caption{The parameter $p/\rho=\omega$ (see Eq. 28 and 31) is
plotted against the time parameter $T$. Other model parameters are
fixed as in Fig.9}
\end{figure}

\newpage
\pagebreak
\section{Conclusion and discussion}

In this section, we discuss the results of our paper. All the
cosmological parameters with the exception of $\omega$ are plotted
in figures in 1 to 12 in logarithmic scale against dimensionless
time parameter $T$. The parameters in section 3.1 are shown in
figures 1 to 4. The cosmological energy density decreases with time
and then remains constant in far future. The cosmological constant
was large in the past which resulted in inflation while now it is
small to produce current accelerated expansion. The Newton's
gravitational constant steadily increased which caused structures to
form. Also the dimensionless parameter $\omega$ varies from the
positive to negative values and converging to $-1$ at current time
$t=t_o(\sim1.2H_o)$ or $T\simeq1$ showing that $\omega$ is
inherently evolving over cosmic history, with $T=0$ corresponds to
the big bang epoch (units are chosen to be meter, kilogram, sec). A
similar behavior is obtained for parameters of section 3.2 and 3.3
shown in figures 5 to 8 and 9 to 12, respectively. In figures (13),
(14) and (15), we have plotted the same parameters against redshift
$z$.

The problem attempted in this paper can also be looked in the
context of bulk viscous cosmology. The anisotropic stresses can be
important at large scale and hence they should be incorporated in
the MCG equation of state (see \cite{mak} for the basic formalism).
It would also be interesting to extend our model using the modified
$f(R)$ gravity theory as well \cite{paul}.

\subsubsection*{Acknowledgment} One of us (MJ) would like to thank
Asghar Qadir for sharing useful comments on this work. We would also
thank the anonymous referee for his useful criticism on this work.

\begin{figure}
\includegraphics{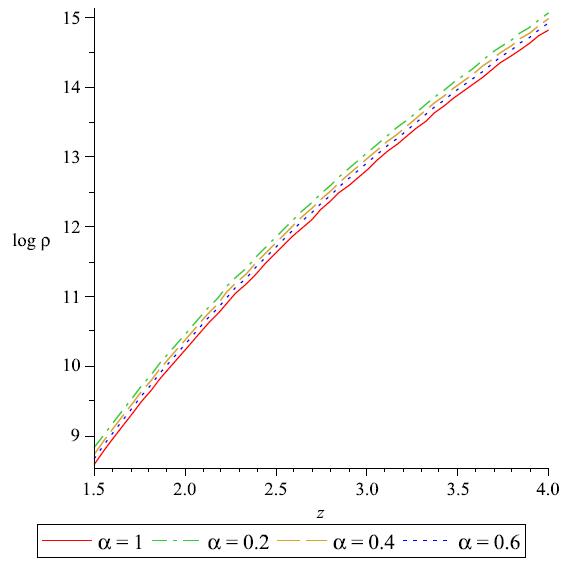}\\
\caption{The energy density $\rho$ (see Eq. 16) is plotted against
redshift parameter $z$. Model parameters are fixed as in Fig. 13}
\end{figure}
\begin{figure}
\includegraphics{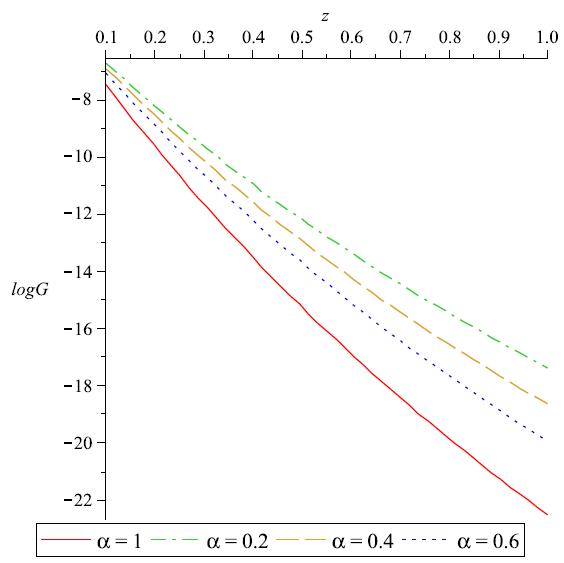}\\
\caption{The Newton's constant $G$ (see Eq. 17) is plotted against
redshift parameter $z$. The model parameters are fixed at $A=B=C=2$,
$\gamma=-0.2$ and $\beta=1$.}
\end{figure}
\begin{figure}
\includegraphics{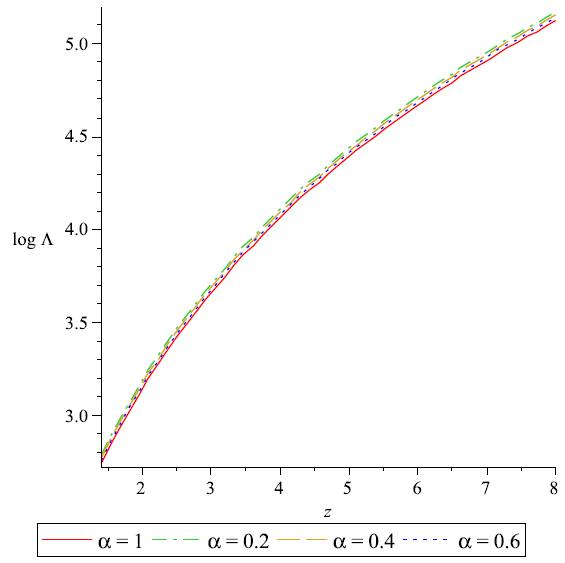}\\
\caption{The cosmological constant $\Lambda$ (see Eq. 18) is plotted
against redshift parameter $z$. Model parameters are fixed as in
Fig. 13}
\end{figure}

\pagebreak
\end{document}